\documentstyle[amssymb,aps,multicol,epsf]{revtex}

\begin{document}
\title{High-field AFMR in single-crystalline La$_{0.95}$Sr$_{0.05}$MnO$_{3}$:
Experimental evidence for the existence of a canted magnetic structure}
\author{A. Pimenov$^{1}$, M. Biberacher$^{1}$, D. Ivannikov$^{1}$, A. Loidl$^{1}$,
V. Yu. Ivanov $^{2}$, A. A. Mukhin $^{2}$, and A. M. Balbashov$^{3}$}
\address{$^{1}$Experimentalphysik V, EKM, Universit\"{a}t Augsburg, 86135
Augsburg, Germany\\
$^{2}$General Physics Institute of the Russian Acad. Sci., 117942 Moscow,
Russia\\
$^{3}$Moscow Power Engineering Institute, 105835 Moscow, Russia}

\maketitle

\begin{abstract}
High-field antiferromagnetic-resonance (AFMR) spectra were obtained in the
frequency range 60 GHz $<\nu <$ 700 GHz and for magnetic fields up to 8 T in
twin-free single crystals of La$_{0.95}$Sr$_{0.05}$MnO$_{3}$. At low
temperatures two antiferromagnetic modes were detected, which reveal
different excitation conditions and magnetic field dependencies. No
splitting of these modes was observed for any orientation of the static
magnetic field excluding the phase-separation scenario for this composition.
Instead, the full data set including the anisotropic magnetization can be
well described using a two-sublattice model of a canted antiferromagnetic
structure.
\end{abstract}


\begin{multicols}{2}

The idea of phase separation in the manganite perovskites (R$_{1-x}$M$_{x}$%
MnO$_{3}$, R=La, Pr ..., M=Ca, Sr ...) is one of the most controversially
discussed topics concerning the electronic properties of these compounds.
After the pioneering works of Jonker and van Santen\cite{jonker} and of
Wollan and Koehler\cite{wollan}, de Gennes\cite{degennes} developed a model
in which the purely antiferromagnetic and insulating LaMnO$_{3}$ on
increasing doping passes through a canted (CAF) ground state and arrives at
a purely ferromagnetic and metallic state at high doping level ($x\gtrsim 0.2
$). This phase diagram was calculated using competing superexchange (SE) and
double exchange (DE)\cite{Zener} interactions.

However, in recent years a number of theoretical models predicted that the
CAF structure becomes unstable against electronic phase separation into
ferromagnetic (FM) and antiferromagnetic (AFM) regions\cite{nagaev,yunoki}.
As discussed by Yunoki et al.\cite{yunoki}, the tendency to the phase
separation seems to be an intrinsic property of the double-exchange model. A
number of experimental data including neutron scattering\cite{hennion} and
NMR\cite{allodi} pointed toward the existence of the phase separation in
different types of manganites. For discussion of the recent results see Refs.%
\cite{yunoki,goodenough}.

It has to be pointed out, that the experimental observation of electronic
phase separation is rather difficult. Already more than forty years ago,
Wollan and Koehler\cite{wollan} stated that, on the basis of neutron
diffraction experiments, it is impossible to decide whether the structure of
the doped manganites is homogeneously canted or
inhomogeneously mixed FM and AFM. Only a few experimental methods
can distinguish between inhomogeneous and homogeneous magnetic phases
because the technique has to be sensitive to the local magnetic structure of
the sample. In addition, a sample quality appears to be of major importance
for these experiments. Antiferromagnetic resonance (AFMR) seems to be an
excellent tool for the solution of the phase-separation problem. The main
parameters of the resonance lines, like position, excitation conditions,
behavior in magnetic field etc., sensitively depend on the local environment
of the magnetic moments. Recently, using this method, we have investigated
the concentration dependence of AFMR-lines in low-doped La$_{1-x}$Sr$_{x}$MnO%
$_{3}$ without external magnetic field\cite{europhys}. The results were
explained within the frame of a two-sublattice model, which strongly
supported the existence of a canted magnetic structure. Most La$_{1-x}$Sr$%
_{x}$MnO$_{3}$ crystals of this series were twinned. However, the samples
with 5\% Sr concentration were identified as untwinned single crystals. This
fact allowed the unambiguous determination of the excitation conditions of
AFMR lines and to carry out detailed investigations in static magnetic
fields. In this paper we present, in addition to the results of the
magnetic-field experiments, anisotropic magnetization curves and compare the
observed data with the predictions of a two-sublattice model. The possible
explanations within phase separation models are also discussed.

La$_{0.95}$Sr$_{0.05}$MnO$_{3}$ single crystals were grown by a floating
zone method with radiation heating\cite{preparation}. X-ray powder
diffraction measurements showed that the crystals were single-phase.
Four-circle X-ray analysis showed the twin-free structure of the crystal.
The temperature dependence of the
dc-resistivity of these samples has been published previously\cite{jetp} and
agrees well with literature data\cite{urushibara}. Plane-parallel plates
of size approximately 8$\times $8$\times $1 mm$^{3}$ were used for
optical measurements. The magnetic measurements were carried out on small
pieces of the same crystals.

The magnetization curves of La$_{0.95}$Sr$_{0.05}$MnO$_{3}$ were measured
using a SQUID magnetometer in fields up to 6.5 T. The transmission spectra
in the frequency range 40 GHz $\leq \nu \leq $ 700 GHz were recorded using a
quasioptical technique utilizing backward-wave oscillators as coherent light
sources\cite{volkov}. Combining this method with a superconducting
split-coil magnet equipped with optical windows allows to carry out
transmission experiments in fields up to 8T. The data were obtained in the
frequency-sweep mode at constant magnetic field. However, in some cases
field-sweep measurements were performed because this procedure enhances the
accuracy of determination of the resonance frequency. The
frequency-dependent transmission spectra were analyzed using the Fresnel
optical formulas for a transmission coefficient of a plane-parallel plate%
\cite{born}. The relative transparency of the sample in the frequency range
investigated resulted in the observation of interference patterns in the
spectra. The observation of these interferences allowed the calculation of
the optical parameters of the sample without measuring the phase shift of
the transmitted signal. The dispersion of the magnetic permeability was
taken into account assuming a harmonic oscillator model for the complex
magnetic permeability:
\begin{equation}
\mu ^{*}(\nu )=\mu _{1}+i\mu _{2}=1+\Delta \mu \nu _{0}^{2}/(\nu
_{0}^{2}-\nu ^{2}+i\nu g)  \label{eqreson}
\end{equation}
where $\nu _{0}$, $\Delta \mu $ and $g$ are eigenfrequency, mode strength
and width of the resonance respectively. The dielectric parameters of the
sample $(n^{*}=n+ik)$ were assumed to behave regular in the vicinity of the
resonance frequency. Hence, the frequency-sweep measurements allowed to
obtain absolute values of the parameters of AFMR lines.

Fig.\,\ref{figmagn} shows the low-temperature magnetization curves of
single-crystalline La$_{0.95}$Sr$_{0.05}$MnO$_{3}$ for different
orientations of the magnetic field. The magnetization along the c-axis shows
a spontaneous magnetization and therefore is identified as the direction of
the weak ferromagnetic moment. The magnetization along the crystallographic
b-axis reveals the weakest field dependence and thus resembles the data of a
simple antiferromagnet along the easy (antiferromagnetic) axis\cite{kittel}.

In order to understand the magnetization measurement and the submillimeter
spectra quantitatively, a two-sublattice model has been adopted. The
two-sublattice model is a widely used approximation to describe the
properties of magnetically ordered materials \cite{moria}. It was originally
applied to the spin-wave spectrum of manganites by de Gennes \cite{degennes}%
. For a realistic description, additional contributions to the free energy,
i.e. the single ion anisotropy $D_{x}\Sigma _{i}S_{xi}^{2}+D_{z}\Sigma
_{i}S_{zi}^{2}$ and the Dzyaloshinsky-Moria (D-M) antisymmetric exchange
interactions $\Sigma _{i,j}{\bf d}_{ij}[S_{i}S_{j}]$ have to be taken into
account. In the classical approximation the free energy at T=0 is given by:

\begin{equation}
\begin{array}{l}
F({\bf m,l})=\frac{1}{2}A{\bf m}^{2}-B|{\bf m}|+\frac{1}{2}%
K_{x}(m_{x}^{2}+l_{x}^{2})+ \\
+\frac{1}{2}K_{z}(m_{z}^{2}+l_{z}^{2})-d(m_{z}l_{y}-m_{y}l_{z})-M_{0}{\bf mH}
\end{array}
\label{eqfree}
\end{equation}

In Eq.\,(\ref{eqfree}) the (x,\thinspace y,\thinspace z) axes of the
coordinate system are directed along the crystallographic axes (a,\thinspace
b,\thinspace c) of the sample (a=5.547\AA ,\thinspace b=5.666\AA ,\thinspace
c=7.725\AA ). The first and the second terms of Eq.\,(\ref{eqfree}) describe
antiferromagnetic and ferromagnetic (double) exchange, the third and fourth
terms give the single ion anisotropy, the fifth term describes the D-M
exchange while the last term takes into account effects of an external
magnetic field. In Eq.(\ref{eqfree}), ${\bf m}$ and ${\bf l}$ are
dimensionless ferro- and antiferromagnetic vectors, which are defined as $%
{\bf m}=({\bf M}_{1}+{\bf M}_{2})/2M_{0}$, ${\bf l}=({\bf M}_{1}-{\bf M}%
_{2})/2M_{0}$ and satisfy the conditions ${\bf ml}=0$, ${\bf m}^{2}+{\bf l}%
^{2}=1$ since the sublattices ${\bf M}_{1}$ and ${\bf M}_{2}$ are assumed to
be saturated at $T=0$. The parameter $B$ describes the DE
interaction. $K_{x,z}>0$ are anisotropy constants stabilizing the $A_{y}F_{z}
$ configuration in pure LaMnO$_{3}$. $d$ is the interlayer antisymmetric
exchange constant. $M_{0}=0.95M_{0}(Mn^{3+})+0.05M_{0}(Mn^{4+})=3.95\mu _{%
\text{B}}$, is the saturation magnetization of the sublattices. The
equilibrium arrangement of the sublattices has been obtained minimizing the
free energy given by Eq.\,(\ref{eqfree}). The frequencies of the resonance
modes were calculated in the limit of small perturbations from the equations
of motion $d{\bf M}_{i}/dt=\gamma [{\bf M}_{i}\times \partial F/\partial
{\bf M}_{i}],$ $(i=1,2)$, where $\gamma $ is the gyromagnetic ratio.

The solutions of Eq.(\ref{eqfree}) for the H%
\mbox{$\vert$}%
\mbox{$\vert$}%
c were published previously \cite{europhys}. The full set of solution for
the field orientation along the a and b-axes is too lengthy and will be
published in full form elsewhere\cite{formeln}. However, in order to
understand the experimental data qualitatively some approximations can
easily be made. Assuming $(B,d,K_{x},K_{z},M_{0}H)\ll A\,$, the approximate
solution for the magnetization can be written as:

\begin{equation}
M_{x}\equiv M_{0}m_{x}=\chi _{\perp }H_{x}(B+d)/d\text{ , }{\bf H}%
=(H_{x},0,0)  \label{eqmx}
\end{equation}

\begin{equation}
M_{y}\equiv M_{0}m_{y}=\chi _{rot}H_{y}\text{ , }{\bf H}=(0,H_{y},0)
\label{eqmy}
\end{equation}

\begin{equation}
M_{z}\equiv M_{0}m_{z}=M_{z}^{0}+\chi _{\perp }H_{z}\text{ , }{\bf H}%
=(0,0,H_{z})  \label{eqmz}
\end{equation}
where $M_{z}^{0}\equiv M_{s}=M_{0}(B+d)/(A+K_{z})$ is the spontaneous
magnetic moment along the c-axis, $\chi _{\perp }=M_{0}^{2}/(A+K_{z})$ and $%
\chi _{rot}=M_{s}^{2}/K_{z}$ are the transverse and rotational
susceptibilities respectively.

The analysis of Eq.\,(\ref{eqmz}) shows that the z-axis exhibits weak
ferromagnetism as the magnetization is nonzero in the absence of an external
magnetic field. The magnetization along the y-axis (Eq. \ref{eqmy}) is
determined by the small rotational susceptibility and disappears in the pure
antiferromagnetic case $(B=d=0)$. The low-field susceptibility along the
x-direction (Eq. \ref{eqmx}) is enhanced compared to the z-axis by the
factor $(B+d)/d$. Qualitatively similar behavior of the magnetization is
observed in Fig. \ref{figmagn}. The solid lines in Fig. \ref{figmagn} were
calculated using the exact expressions based on Eq.\,(\ref{eqfree}) and
describe the experimental data reasonably well. A small static moment along
the y-axis appears to be strongly angle dependent and possibly is due to
some residual influence of the spontaneous moment along the z-axis. The
absolute values of the parameters of the model were obtained by
simultaneously fitting the magnetization curves and the values of the
resonance frequencies in the absence of magnetic field. Despite the
relatively large number of parameters in Eq.\,(\ref{eqfree}) $%
(A,B,d,K_{x},K_{z})$, the requirement of a simultaneous fit allows the
unambiguous determination of the parameters: $A=4.67\cdot 10^{7}$ erg/g, $%
B=7.4\cdot 10^{6}$ erg/g, $K_{z}=3.33\cdot 10^{6}$ erg/g, $K_{x}=3.42\cdot
10^{6}$ erg/g, $d=2.1\cdot 10^{6}$ erg/g, and $M_{0}=92.14$ emu/g. From the
values of $A$ and $K_{x}$ the interlayer exchange ($%
J_{2}=-0.37meV$) and the single-ion anisotropy ($C=0.11meV$) constants can be
calculated\cite{europhys} which are in good agreement with neutron
scattering data for La$_{0.95}$Ca$_{0.05}$MnO$_{3}$\cite{laca05} and for La$%
_{0.95}$Sr$_{0.06}$MnO$_{3}$\cite{hennion}.

Fig. \ref{figspectra} shows the transmission spectra of the La$_{0.95}$Sr$%
_{0.05}$MnO$_{3}$ single crystal at low temperatures. The solid lines were
calculated using the Fresnel equations and Eq.(\ref{eqreson}) as described
above. The parameters of the magnetic mode were obtained by fitting the
transmission spectra. In addition, the frequency position and the appearance
of the modes were also examined using field sweeps at fixed frequencies. From
the data shown in Fig.\,\ref{figspectra} two peculiarities of the
observed AFMR modes immediately become clear: i) the observed lines have
unique excitation conditions: $\widetilde{h}\parallel $ c -axis for the
high-frequency mode and $\widetilde{h}\parallel $ b -axis for the
low-frequency mode, and ii) no splitting of the AFMR lines is observed in
finite magnetic fields for any geometry of the experiment. Both conclusions
are characteristic properties of a canted antiferromagnetic structure\cite
{herrmann} and follow naturally from the solution of Eq.\,(\ref{eqfree}).

The magnetic field dependencies of the resonance frequencies of both AFMR
lines are shown in Fig.\,\ref{figmode}. The solid lines in Fig.\,\ref{figmode}
were calculated on the basis of the two-lattice model, discussed above.
However, the parameters of the model {\em were already fixed} by fitting the
magnetization curves and absolute values of the AFMR frequencies in the
absence of magnetic field. Having this in mind, the theoretical curves
describe the experimental data reasonably well. The most important feature
of Fig. \ref{figmode} is the softening of the FM-mode for B$\parallel $b.
This softening represents a common property of magnetic resonance in
antiferromagnets and is followed by the field-induced rearrangement of the
magnetic structure (spin-flop) at a critical value of magnetic field. The
softening of the FM-mode at low fields is in good agreement with the model
calculations. However, the behavior for higher fields (B$\sim $7T)
significantly deviates from the model predictions. These deviations are most
probably due to the extreme sensitivity of the data with respect to the
exact orientation of the static magnetic field and the neglect of the
higher-order terms in Eq.\,(\ref{eqfree}). The angular dependence of a
critical behavior in a canted antiferromagnet has been calculated in details
by Hagedorn and Gyorgy \cite{angledep}. These calculations show that already
a misalignment of the magnetic field as low as one degree strongly suppress
the softening of the FM-line in the vicinity of the critical field. Most
probably similar effects explain the deviations observed in Fig. \ref
{figmode}.

Within the presented model it is also possible to calculate the absolute
intensities of the AFMR modes. In the absence of static field the solution
of Eq. (\ref{eqfree}) gives $\Delta \mu _{xx}=0.0080$ and $\Delta \mu
_{zz}=0.0136$, using the parameters obtained above. These values are in good
agreement with the experimental values $\Delta \mu _{xx}=0.012\pm 0.003$ and
$\Delta \mu _{zz}=0.0120\pm 0.0010$.

Finally, we discuss the possible explanation of the above-presented data
within the concept of phase separation in the ferromagnetic droplets in an
antiferromagnetic matrix. Already the magnetization data (Fig. \ref{figmagn}%
) impose a set of constraints on the possible configuration of the phases.
E.g. ferromagnetic moments have to be parallel to the c-axis and the b-axis
has to be the antiferromagnetic easy axis. However, the most important
consequences of a possible electronic phase separation follow for the
properties of the magnetic resonances:

- the antiferromagnetic phase would reveal an AFMR\ mode with a resonance
frequency similar to the frequencies in pure LaMnO$_{3}$ ($\nu \sim $18 cm$%
^{-1}$).

- this AFMR mode should split into two modes in the presence of magnetic
field, as was observed by Mitsudo et al.\cite{mitsido} in pure LaMnO$_{3}$.

- a ferromagnetic line arising from the ferromagnetic droplets has to be
observed. The frequency of this mode is expected at substantially lower
frequencies ($\nu \sim 10GHz$) as observed by Lofland et al.\cite{lofland}
in La$_{0.9}$Sr$_{0.1}$MnO$_{3}$. The resonance frequency of this mode
should increase roughly linear with external magnetic field up to
frequencies 150-250 GHz for B=7T.

None of these properties could be detected in the present experiment.
Instead, the observed picture can be well described using the canted
magnetic structure.

In conclusion, twin-free single crystals of La$_{0.95}$Sr$_{0.05}$MnO$_{3}$
were grown by the floating-zone method. The low-temperature magnetization
was measured along the principal crystallographic directions. High field
AFMR spectra of this sample were investigated in the frequency range 60-700
GHz and for magnetic fields up to B=8T. Two AFMR lines having different
excitation conditions were detected at low temperatures. The softening of
the low-frequency mode was observed for the orientation of the static field B%
\mbox{$\vert$}%
\mbox{$\vert$}%
b and is explained by approaching to the spin-flop transition. The full data
set, obtained for the La$_{0.95}$Sr$_{0.05}$MnO$_{3}$ single crystal, can be
easily explained as arising from a canted magnetic structure and clearly
contradicts the concept of phase separation into ferro- and
antiferromagnetic regions.

This work was supported in part by BMBF (13N6917/0 - EKM), by DFG (Pi
372/1-1), by RFBR (99-02-16849), and by INTAS (97-30850).

\begin{figure}[tbp]
\caption{ Magnetization of La$_{0.95}$Sr$_{0.05}$MnO$_{3}$ single crystal
along different crystallographic axes at T=4.2K. Symbols - experiment, lines
are calculated according to the two-lattice model as described in the text. }
\label{figmagn}
\end{figure}

\begin{figure}[tbp]
\caption{ Submillimeter-wave transmission spectra of single crystalline
 La$_{0.95}$Sr$_{0.05}$MnO$_{3}$ for two different geometries of the experiment
at low temperatures. The lines were calculated using the Fresnel equations
for the trassmission of a plane-parallel plate. The behavior of the
complex susceptibility was described by Eq.\,(\ref{eqreson}). }
\label{figspectra}
\end{figure}

\begin{figure}[tbp]
\caption{ Magnetic field-dependence of the resonance frequencies of the AFMR
lines in La$_{0.95}$Sr$_{0.05}$MnO$_{3}$ at low temperatures.
Solid lines were calculated as described in the text.
}
\label{figmode}
\end{figure}

\end{multicols}

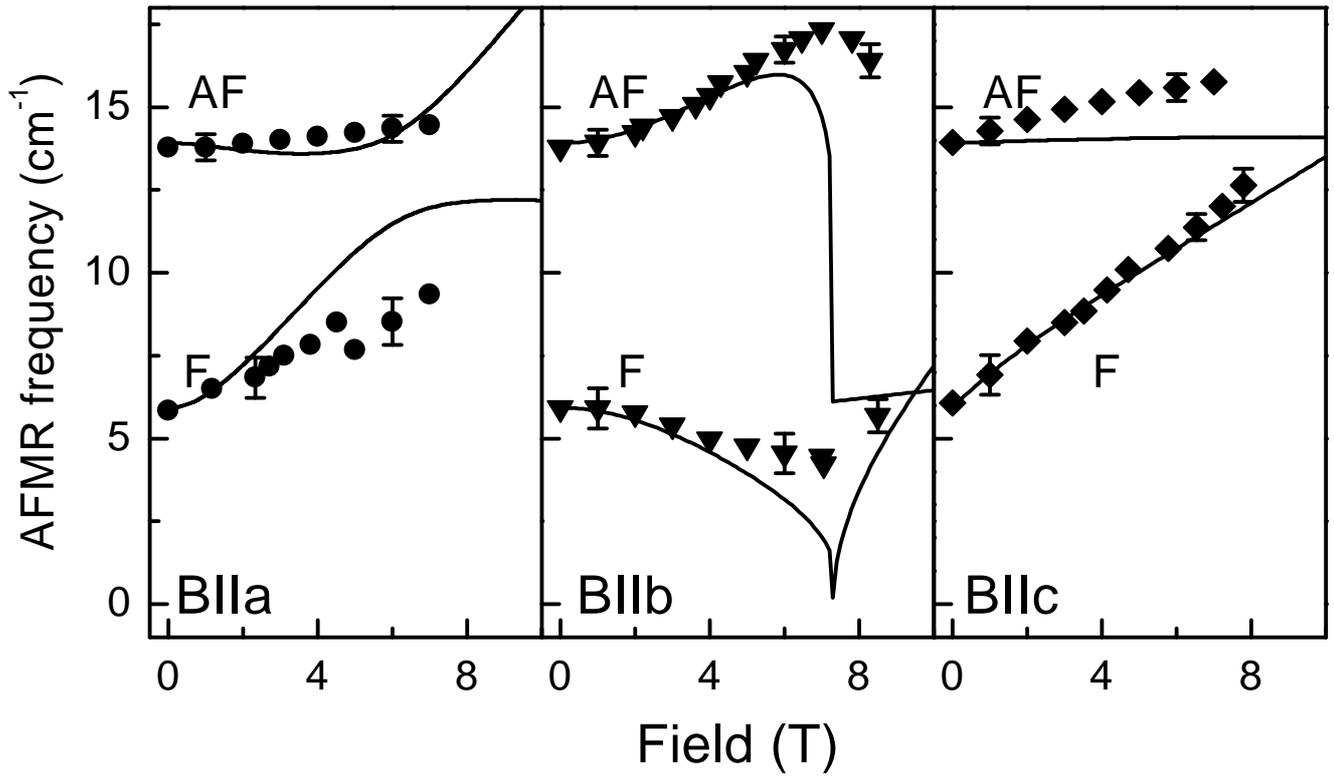
\end{document}